\documentclass[journal,letterpaper]{IEEEtran}

\usepackage{graphicx,cite,epsfig,amssymb,amsmath,amsfonts,multicol,subfigure,mathtools,bm,mathrsfs,setspace}
\usepackage{multirow}
\usepackage{array}

\newcommand{\tabincell}[2]{\begin{tabular}{@{}#1@{}}#2\end{tabular}}

\begin{document}
\title{\huge Beyond the Channel Capacity of BPSK Input}
\author{
	
	Bingli~Jiao,
	Dongsheng Zheng,
Mingxi Yin and
Yuli Yang
\thanks{B. Jiao ({\em corresponding author}) is with the Department of Electronics and Peking University-Princeton University Joint Laboratory of Advanced Communications Research, Peking University, Beijing 100871, China (email: jiaobl@pku.edu.cn).}
\thanks{Dongsheng Zheng is with the Department of Electronics and Peking University-Princeton University Joint Laboratory of Advanced Communications Research, Peking University, Beijing 100871, China (email: zhengds@pku.edu.cn).}
\thanks{Mingxi Yin is with the Department of Electronics and Peking University-Princeton University Joint Laboratory of Advanced Communications Research, Peking University, Beijing 100871, China (email: yinmx@pku.edu.cn).}
\thanks{Y. Yang is with the Department of Electronic and Electrical Engineering, University of Chester, Chester CH2 4NU, U.K. (e-mail: y.yang@chester.ac.uk).}

}

\maketitle
   
\begin{abstract} 
The paper proposed a method that organizes a parallel transmission of two signals to be separated from each other at receiver through Hamming- to Euclidean space, where the conventional problem of achievable bit rate (ABR) is converted to that of the ABR's summation with respect to the two separated signals. Actually, the mutual information is separated accordingly into two. Since the mutual information is non-linear function of signal power in general, the separation brings a chance for achieving higher ABR. The proposed work proceeds along the above thinking and achieves higher bit-error-rate (BER) performance in comparison with BPSK as shown in the simulations. Moreover, one can find theoretically beyond of channel capacity of BPSK input.     
\end{abstract}   
     

\IEEEpeerreviewmaketitle

\section{Introduction}

The channel capacities of finite alphabets can be predicted by the theoretical calculations of the mutual informations and confirmed by bit error rate (BER) performances with the coded modulation schemes.  So far, the smallest gap between the theoretical result and the coding performance is found at 0.0045dB using the low density parity-check code (LDPC) with BPSK input \cite{dB00045}.  

A standard way of modeling communication has been set up with the additive white Gaussian noise (AWGN) channel  
\begin{eqnarray}
\begin{array}{l}\label{equ-ch}
y = x + n,
\end{array}
\end{eqnarray}
where $y$ is the received signal, $x$ is the transmitted signal and $n$ is the received AWGN component from a normally distributed ensemble of power $\sigma_N^2$, denoted by $n \sim \mathcal{N}(0,\sigma_N^2)$.  

At finite-alphabet input, channel capacity can be expressed in terms of the achievable bit rates (ABRs) by the mutual information as  
\begin{equation}
\begin{array}{l}\label{equ1}
\rm{I}(X;Y) =\rm{H}(Y) - \rm{H}(N),
\end{array}
\end{equation}
where $\rm{I}(X;Y)$ is the mutual information, ${\rm{H}}(Y)$ is the entropy of the received signal and ${\rm{H}}(N) = {\log _2} (\sqrt{2 \pi e \sigma_N^2})$ is the entropy of the AWGN \cite{Shannon1948}. 

Our proposed transmission strategy is inspired by a consideration on the signal separation at \eqref{equ-ch} and consequently \eqref{equ1} \cite{jiao}.  Let us envision that the input signal $x$ can be separated into $x_1$ and $x_2$ at the receiver and, subsequently, \eqref{equ-ch} can be re-written as   
\begin{eqnarray}
\begin{array}{l}\label{s-1}
y_1 = x_1 + n,
\end{array}
\end{eqnarray}
and 
\begin{eqnarray}
\begin{array}{l}\label{s-2}
y_2 = x_2 + n,
\end{array}
\end{eqnarray}
where $x_1$ and $x_2$ are the separated signals, and $y_1$ and $y_2$ are the received signals accordingly.  

As such, instead of \eqref{equ1}, the ABR of the system with these separated signals will be calculated using
\begin{eqnarray}
\begin{array}{l}\label{mutual}
\tilde{I}_t=\tilde{I}_{x_1}(E_{x_1}/\sigma_N^2)+\tilde{I}_{x_2}(E_{x_2}/\sigma_N^2)
\end{array}
\end{eqnarray}
where $\tilde{I}_t$ and $\tilde{I}_{x_i}(E_{x_i}/\sigma_N^2)$ are the overall mutual information and the mutual information pertaining to the transmissions of $x_i$ for $i=1,2$, with the argument of the signal-to-noise power ratio (SNR).  Moreover, $E_{x_i}$ is the energy of $x_i$.

Since the result of \eqref{mutual} is not necessarily equal to \eqref{equ1}, one can have a different approach for achieving higher ABR.  Actually,  this work goes through \eqref{s-1} and \eqref{s-2} aiming to improve the capacity achieved by BPSK input, i.e., to reduce the gap of 0.0045dB.

Throughout this paper, we use the capital letter to denote a vector in Hamming space and the lowercase to indicate its components, e.g., $A = \{a_1,a_2, ...., a_M\}$, where $A$ represents the vector and $a_i$ is the $i$th component.  We use bold face letters to denote signals in Euclidean space with two dimensional complex plane.  In the derivations, we use $\hat{y}$ to express the estimate of $y$ at the receiver and $\tilde{\rm{I}}_x(\lambda)$ to express the mutual information $\rm{I}(X;Y)$ with the averaged SNR $\lambda$ as the argument \cite{Verdu2007}.
 
\section{Transmission Scheme}
Consider two independent binary source bit sequences, expressed
in a vector form of $C^{(i)} = \{c^{(i)}_1,c^{(i)}_2,..,c^{(i)}_{k_i},...,c^{(i)}_{K_i}\}$, where  $c^{(i)}_{k_i}$ is the $kth$ information bit (info-bit) of $C^{(i)}$ and $K_i$ is the length of the source subsequence, and $i=1,2$ indicates the two source bit sequences, respectively. 

These two source bit sequences are encoded into the two channel code words expressed as $V^{(i)} = \{v^{(i)}_1,v^{(i)}_2,..,v^{(i)}_m, ...,v^{(i)}_{M}\}$ of equal length $M$ by using two different channel coding matrices
\begin{eqnarray}
\begin{array}{l}\label{equ-code-1}
v^{(i)}_{m} = \sum\limits_{k}g^{(i)}_{{m}k}c^{(i)}_{k}  
\end{array}
\end{eqnarray}  
where $v^{(i)}_m \in \{1,0\}$ is the $m$th code-bit of $V^{(i)}$, and $g^{(i)}_{mk}$ is the element of the code matrix $G^{(i)}$, $m=1,2,....,M$, where $M$ is length of the channel code word. 

The code-bits of $V^{(1)}$ and $V^{(2)}$ are mapped onto two signal sets represented by $x_1$ and $x_2$ in \eqref{s-1} and \eqref{s-2}, respectively.  For simplifying expressions of the signal organization, we omit the indices of the code-bits series of $V^{(1)}$, $V^{(2)}$ and, thus, work on the modulation of $v^{(1)}$ and $v^{(2)}$ without losing generality in the following derivations.  

First, we define $x_1$ by the conventional BPSK in the complex plane to modulate each code-bit in $V^{(1)}$   
\begin{eqnarray}
\textbf{x}_1 = \left\{
\begin{array}{l}\label{x_1}
(\sqrt{E_s},j0),    \ \ \ \ \  \ \ \ v^{(1)}=0,  \\
(-\sqrt{E_s}, j0),   \ \ \ \  \ \ v^{(1)}=1,   
\end{array}
\right.
\end{eqnarray}  
with $j=\sqrt{-1}$, where $E_s$ is the symbol energy of $\textbf{x}_1$.

For constructing $x_2$, we define a rotation operator $\Gamma_{\beta}=e^{j\beta}$ that rotates a vector for an angle $\beta$ in complex plane, e.g., 
\begin{eqnarray}
\begin{array}{l}\label{x_2}
\textbf{z}'= \Gamma_{\beta}\textbf{z} =\textbf{z}e^{j\beta}
\end{array}
\end{eqnarray}
where $\textbf{z}'$ is the vector from the rotation of $\textbf{z}$ by angle $\beta$. To complete the operator, we note that $\Gamma_0=1$ and $\Gamma_{\beta}\Gamma_{-\beta}=1$ hold. 

Each code-bit of $V^{(2)}$ is mapped onto $\beta$ as the follows:  $v^{(2)}=0$ is mapped onto $\beta=0$ and $v^{(2)} =1$ is mapped onto $\beta =\pi/2 $.   The signal $x_2$ is defined in a vector form of $\textbf{x}_2$ in the complex plane as  
\begin{eqnarray}
\begin{array}{l}\label{x_2}
\textbf{x}_2= \Gamma_{\beta}\textbf{x}_1 =\textbf{x}_1e^{j\beta},
\end{array}
\end{eqnarray}
where $\textbf{x}_2$ is the rotated $\textbf{x}_1$ with an angle $\beta$.

\begin{table}[htb]
	\renewcommand{\arraystretch}{1.5}
	\centering
	\small
	\caption{Signal modulation results.}
	\label{Table1}
	\begin{tabular}{|p{2cm}<{\centering}|p{2cm}<{\centering}|p{2cm}<{\centering}|}
		\hline
		\tabincell{c}{$v^{(2)}$} & \tabincell{c}{$\beta$} & \tabincell{c}{$x_{2}$} \\
		\hline
		\multirow{2}{*}{$0$} & \multirow{2}{*}{$0$} & \tabincell{c}{$s_{1}$} \\
		\cline{3-3}
		& & \tabincell{c}{$s_{3}$} \\
		\hline
		\multirow{2}{*}{$1$} & \multirow{2}{*}{$\pi/2$} & \tabincell{c}{$s_{2}$} \\
		\cline{3-3}
		& &\tabincell{c}{$s_{4}$} \\	
		\hline
		
	\end{tabular}
\end{table}    

Since $\textbf{x}_1$ has two possible values, each value of $v^{(2)}$ can be mapped onto two possible points in the complex plane as shown in Fig.1, where
$\textbf{s}_1$ and $\textbf{s}_3$ are the mapping results of $v^{(2)}=0$, and $\textbf{s}_2$ and $\textbf{s}_4$ the results of $v^{(2)}=1$.  We refer this method to as the double mapping modulation (DMM), because one value of the code-bit is mapped onto two points in Euclidean space.  The DMM is listed in Table I for the use of demodulation latter. 

It is noted that the symbol energy of $\textbf{x}_2$ is also equal to $E_s$.

After the above modulations, we can find that $\textbf{x}_2 \in \{\textbf{s}_1, \textbf{s}_2, \textbf{s}_3, \textbf{s}_4\}$ and the symbol $\textbf{s}_k$, for $k = 1, 2, 3, 4$, contains one bit from $V^{(1)}$ and another bit from $V^{(2)}$.  

In the proposed method, the transmitter inputs $\textbf{s}_k$ to the AWGN channel
\begin{eqnarray}
\begin{array}{l}\label{equ-code}
\textbf{y}=\textbf{s}_k + n = \Gamma_{\beta}\textbf{x}_1 + \textbf{n}
\end{array}
\end{eqnarray}  
where $\textbf{n}$ is the noise in a vector form.    

\begin{figure}[htb]
	\centering
	\includegraphics[width=0.3\textwidth]{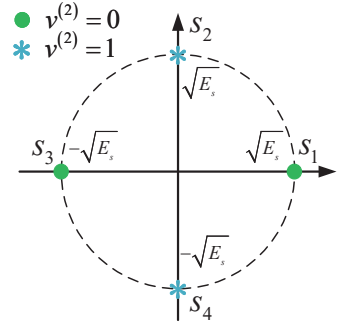}
	\caption{Constellation of the proposed scheme.}
	\label{Fig1}
\end{figure}

At the receiver, all the signals are recoded in a storage because each of the signals will be used twice: one time for demodulation of $V^{(2)}$ and another time for $V^{(1)}$.

The information recoveries are done through the inverse steps of the transmitter with demodulation of $V^{(2)}$ first  
\begin{equation}
v^{(2)} = \left\{
\begin{array}{l}\label{equ3}
0, \qquad \hat{\textbf{y}} = \textbf{s}_1 \ \ or \ \ \textbf{s}_3,   \\
1, \qquad \hat{\textbf{y}} =\textbf{s}_2 \ \ or \ \ \textbf{s}_4,
\end{array}
\right.
\end{equation}
where $\hat{\textbf{y}}$ is the estimate of $\textbf{y}$ for demodulation of $v^{(2)}$. Then $\hat{C}^{(2)}$ is recovered using the conventional decoding scheme. 

Once $\hat{C}^{(2)}$ has been obtained, the receiver will find each value of $\beta$ by reconstructing $V^{(2)}$ with the code matrix of \eqref{equ-code-1}.  Then, by using table I, one can find the rotation angle $\beta$ and recover the BPSK modulation of $\textbf{x}_1$ by 
\begin{eqnarray}
\begin{array}{l}\label{equ-recove}
\hat{\textbf{y}}_1= \Gamma_{-\hat{\beta}\beta}\textbf{y} = \hat{\textbf{x}}_1 + \textbf{n}'
\end{array} 
\end{eqnarray}
where $\hat{\textbf{y}}_1$ is the recovered BPSK symbol plus the noise, $\textbf{y}$ is the reused signal and $\hat{\beta}$ is the estimated rotation angle obtained by using Table 1.

Finally, one can decode with the results of \eqref{equ-recove} and obtain the recovered $\hat{C}^{(1)}$.  

Since the rotation operation does not change the amplitudes of both signal- and noise vector, the SNR in \eqref{equ-recove} remains unchanged as compared to that in \eqref{equ-code}.

\section{Beyond Capacity of BPSK Input}
In this section, we work on the BER simulations to show the advantage of the proposed signal transmission scheme.

For a given symbol energy $E_s$, the signal $\textbf{x}_1$ can achieve exactly same BER performance as that of BPSK input when $\hat{\beta}$ is of error free in \eqref{equ-recove}.  

Theoretically, according to Shannon theorem, there exist a channel code of infinitive length that can reduce the error probability of the info-bits with $ \hat{C}^{(2)} $ to infinitive small.  Consequently, the error probability of recovered code-bits $\hat{V}^{(2)}$ can be also of arbitrary small because the linear relation between the source- and channel code.  Thus,  the error free assumption of $\hat{\beta}$ is correctly.   

Then, the ABR contribution from $\textbf{x}_2$ increases the overall ABR of the proposed method to a level beyond the channel capacity of conventional BPSK.

From practical point of view, one can find the gap between ABR of BPSK and the channel capacity is at 0.0045dB[1]. Then, we need to find a net ABR contribution from $\textbf{x}_2$ in case error presents in $\hat{\beta}$ pertaining to errors in  $\hat{V}^{(2)}$. 

Let us study the degradation of $\text{x}_1$ by simulating the signal communication scheme of section II exactly.  Since the target spectral efficiency is at 0.5bit/Hz/s, we use the LDPC code (DVB-S.2) of bits' length of 68000 with code rate 1/2 to work with $\textbf{x}_1$ and the LDPC of code rates of 1/8 and 1/16 constructed by repeating each code-bit of the LDPC of code rate 1/4 for K times, where K=2 and 4.  

Aiming at BER of $10^{-8}$, we examine how many dB loss of the BPSK with $\textbf{x}_1$ due to the errors in $\hat{V}^{(2)}$.  The simulation results are shown in Fig. 2, where one can find the degradations of 0.1dB and 0.7dB loss in terms of symbol energy to noise ratios, with respect to K= 2 and 4, i.e. code rates of 1/8 and 1/16, respectively.   

\begin{figure}[htb]
	\centering
	\includegraphics[width=0.45\textwidth]{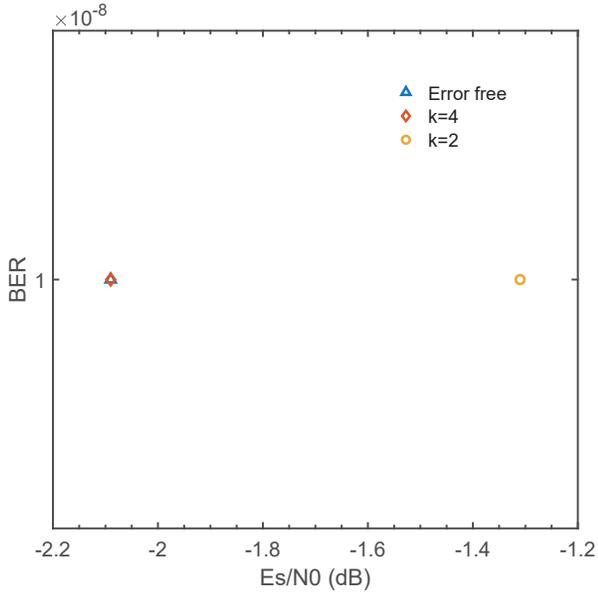}
	\caption{Degradation of K=2 and 4, i.e., code rate equal 1/8 and 1/16 respectively}
	\label{fig4}
\end{figure}

\begin{figure}[htb]
	\centering
	\includegraphics[width=0.45\textwidth]{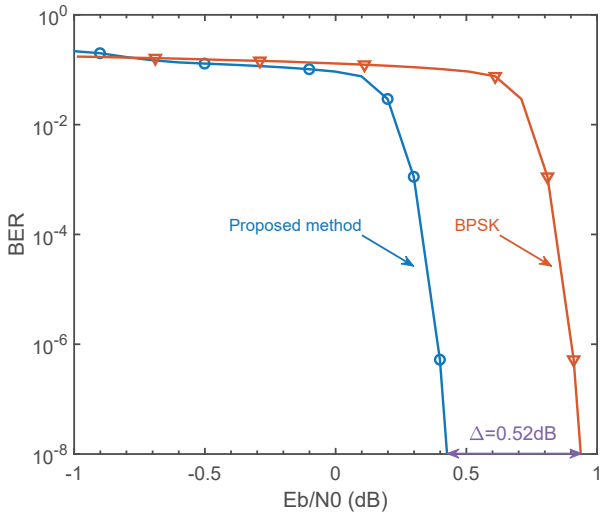}
	\caption{The difference between the proposed scheme and QPSK.}
	\label{fig5}
\end{figure}

Then, we select the constructed LDPC of code rate 1/16 to $\textbf{x}_2$ and the LDPC of code rate 1/2 to $\textbf{x}_1$ to simulate the scheme of section II, compare with conventional BPSK plus the same LDPC used by $\textbf{x}_1$ and find eventually 0.052dB gain as shown in Fig.3.     

An extension is made by using the 0.52dB gain to add to 0.0045dB of \cite{dB00045}, where one can find that the gap between the extension of this approach and channel capacity can be changed to  G=$0.0045dB-0.52dB =-0.516dB$ that indicates the beyond of the channel capacity of BPSK input, where G is the new gap obtained by the extension.     

It is noted that the extension showing beyond the channel capacity is a conservative estimation because using two best existed LDPC codes to both $\textbf{x}_1$ and $\textbf{x}_2$ can reach the largest gain, while the extension is made using only one to $\textbf{x}_1$.  The constructed code of code rate 1/16 can be far from the best existed code at the BER performance for helping the overall performance of this approach.

\section{Conclusion}

In this paper, we proposed the parallel transmission method that can be separated at the receiver, where the signal separation is enabled by our created DMM working through Hamming- and Euclidean space.  The simulation results show the better BER performance in comparison with conventional BPSK and both the theoretical analysis and the extension of this approach show the beyond channel capacity of BPSK input.


\begin{thebibliography}{1}

	\bibitem{dB00045}
	Sae-Young Chung, G. D. Forney, T. J. Richardson and R. Urbanke, ``On the design of low-density parity-check codes within 0.0045 dB of the Shannon limit," in IEEE Communications Letters, vol. 5, no. 2, pp. 58-60, Feb 2001.
	
  \bibitem{Shannon1948}
	C. E. Shannon, ``A mathematical theory of communication'', \textit{The Bell System Technical Journal}, vol. 27, no. 3, pp. 379-423, July 1948.
	
	\bibitem{jiao}
	B. Jiao and D. Li, ``Double-space-cooperation method for increasing channel capacity'', \textit{China Communications}, vol. 12, no. 12, pp. 76-83, Dec. 2015.

	
	\bibitem{Verdu2007}
	D. P. Palomar and S. Verdú, ``Representation of Mutual Information Via Input Estimates'', \textit{IEEE Trans. Inform. Theory}, vol. 53, no. 2, pp.453-470, 2007.
	

	\bibitem{Sayana2008}
	K. Sayana, J. Zhuang and K. Stewart, "Short Term Link Performance Modeling for ML Receivers with Mutual Information per Bit Metrics," IEEE GLOBECOM 2008 - 2008 IEEE Global 	  Telecommunications Conference, New Orleans, LO, 2008, pp. 1-6.
	
	
	
\end{thebibliography}
\end{document}